\documentstyle{mn}

\newcommand{\reference}{\bibitem}
\def\mnras{MNRAS}
\def\araa{ARAA}
\def\aap{A\&A}
\def\apj{ApJ}
\def\plotone#1{\centering \leavevmode
\epsfxsize=\columnwidth \epsfbox{#1}}

\def\beq{\begin{equation}}
\def\eeq{\end{equation}}
\def\bey{\begin{eqnarray}}
\def\eey{\end{eqnarray}}
\def\kms{{\rm \,km\,s^{-1}}}

\def\chisq{\chi^2}

\def\tE{t_{\rm E}}
\def\rEt{\tilde{r}_{\rm E}}

\def\u0{u_0}

\def\Ds{D_{\rm s}}
\def\Dl{D_{\rm l}}
\def\mI0{m_{I,0}}

\def\fl{f_{\rm L}}

\def\pirel{\pi_{\rm rel}}
\def\snpeak{({\rm S/N})_{\rm peak}}

\newcommand{\up}[1]{\ifmmode^{\rm #1}\else$^{\rm #1}$\fi}

\newcommand{\arcd}{\ifmmode^{\circ}\else$^{\circ}$\fi}
\newcommand{\arcm}{\ifmmode{'}\else$'$\fi}
\newcommand{\arcs}{\ifmmode{''}\else$''$\fi}

\input epsf
\begin{document}

\title[Parallax Microlensing Events in the OGLE II Database Toward 
the Galactic Bulge]
{Parallax Microlensing Events in the OGLE II Database Toward 
the Galactic Bulge}

\author[Smith, Mao \&  Wo\'zniak]
{Martin C. Smith$^1$, Shude Mao$^1$, P. Wo\'zniak$^{2,3}$
\thanks{e-mail: (msmith, smao)@jb.man.ac.uk, wozniak@lanl.gov
}
\thanks{
Based on observations obtained with the 1.3 m Warsaw
Telescope at the Las Campanas Observatory of the Carnegie
Institution of Washington.
}
\\
\smallskip
$^{1}$Univ. of Manchester, Jodrell Bank Observatory, Macclesfield,
Cheshire SK11 9DL, UK \\ 
$^{2}$Princeton University Observatory, Princeton, NJ 08544-1001, USA \\
$^{3}$Los Alamos National Laboratory, MS D436, Los Alamos,
NM 87545, USA
}
\date{Accepted ........
      Received .......;
      in original form ......}

\pubyear{2001}

\maketitle
\begin{abstract}
We present a systematic search for parallax microlensing events 
among a total of 512 microlensing candidates
in the OGLE II database for the 1997-1999 seasons.
We fit each microlensing candidate
with both the standard microlensing model and also a parallax model that
accounts for the Earth's motion around the Sun.  We then 
search for the parallax signature by comparing the $\chi^2$ of the
standard and parallax models. For the events which show a
significant improvement, we further use the `duration' of the 
event and the signal-to-noise ratio as criteria to 
separate true parallax events from other noisy microlensing events. 
We have discovered one convincing new candidate, sc33\_4505, 
and seven other marginal cases. The convincing candidate 
(sc33\_4505) is caused by a
slow-moving, and likely low-mass, object, similar to other known
parallax events. We found
that irregular sampling and gaps between observing seasons
hamper the recovery of parallax events.
We have also searched for
long-duration events that do not show parallax signatures. The lack
of parallax effects in a microlensing event puts a lower-limit on 
the Einstein radius projected onto the observer plane, which in turn 
imposes a lower limit on the 
lens mass divided by the relative lens-source parallax.
Most of the constraints are however quite weak.
\end{abstract}


\section{Introduction}

Gravitational microlensing was originally proposed as a method of detecting
compact dark matter objects in the Galactic halo (Paczy\'nski
1986). However, it also turned out to be an extremely useful method to
study Galactic structure, mass functions of stars and potentially extra-solar
planetary systems (for a review, see Paczy\'nski 1996). 
Most microlensing events are well described by the standard light curve
(e.g. Paczy\'nski 1986). Unfortunately, from these light curves, one
can only derive a single physical constraint, namely the Einstein radius
crossing time, which involves the lens mass, various distance measures
and relative velocity (see \S 3). This degeneracy means that the lens
properties cannot be uniquely inferred, thus making the
interpretation of the microlensing results ambiguous; this ambiguity
is particularly serious for the interpretation of microlensing events
toward the Large Magellanic Clouds (e.g. Sahu 1994; 
Zhao \& Evans 2000; Alcock et al. 2000b, and references therein).

Fortunately, some microlensing events are exotic, in that they deviate from
the standard shape appropriate for a point source lensed by a
single star (e.g. Paczy\'nski 1986). The
deviations include binary microlensing events
(e.g., Mao \& Paczy\'nski 1991;  Udalski et al. 2000; Alcock et
al. 2000a), the finite source size events 
(Gould 1994; Nemiroff \& Wickramasinghe 1994; Witt \& Mao 1994)
and parallax microlensing events. The importance of these non-standard events
is that they allow one to derive extra constraints on the lens parameters.
Parallax signatures in microlensing events arise when the
event lasts long enough that the Earth's motion can no longer be
approximated as rectilinear during the event (Gould 1992;
see also Refsdal 1966).
Unlike the light curves for the standard events which are symmetric,
these parallax events exhibit asymmetries in their light curves due 
to this motion of the Earth around the Sun. These events
allow one to derive the physical dimension of the
Einstein radius projected onto the observer plane (i.e., the solar system).
The first parallax microlensing event was reported 
by the MACHO collaboration toward the Galactic bulge (Alcock et al. 1995),
and the second case (toward Carina) was discovered by the OGLE collaboration
and reported in Mao (1999). Additional parallax microlensing
candidates have been presented in a conference abstract
(Bennett et al.\ 1997). A spectacular microlensing event
that exhibits parallax signatures over a span of two years was
reported by Soszy\'nski et al. (2001). The MOA collaboration
also discovered a parallax microlensing event toward the
Galactic bulge (Bond et al. 2001). However, despite the importance
 of parallax events, there have been no 
reported systematic searches in the existing databases.
This paper is an attempt to remedy this situation. We search
for parallax signatures among 512 microlensing candidates
discovered by Wo\'zniak et al. (2001) using the difference image analysis. 
The outline of the paper is as follows. In \S2 we briefly describe
the microlensing database, in \S3 we describe our fitting and
search procedures, and in \S4 we describe our candidate parallax events,
including both convincing and marginal cases. We also search the
long-duration events that show no obvious parallax signatures and study
their constraints on the lens parameters. Finally,
in \S5 we discuss the implications of our results and 
highlight observational issues in identifying parallax events.

\section{Database of Microlensing Events}

All observations presented in this paper were carried out during the
second phase of the OGLE experiment with the 1.3 m Warsaw telescope
at the Las Campanas Observatory, Chile. The observatory is operated
by the Carnegie Institution of Washington. The telescope was
equipped with the ``first generation'' camera with a SITe 3
2048$\times$2048 pixel CCD detector working in the drift-scan mode.
The pixel size was 24$\mu$m, giving the scale of 0.417\arcs pixel$^{-1}$.
Observations of the Galactic bulge fields were performed in the
``medium'' speed reading mode with the gain 7.1 e$^{-}$ ADU$^{-1}$
and readout noise about 6.3 e$^{-}$. Details of the instrumentation
setup can be found in Udalski, Kubiak \& Szyma\'nski (1997).
The majority of the OGLE-II frames were taken in the $I$-band,
roughly 200-300 frames per field during observing seasons 1997--1999.
A typical observing season for the Galactic bulge lasts between
mid February to the end of October, which unfortunately produces
gaps of more than a quarter of a year in length (for more details,
see Section~5).
Udalski et al. (2000) gives full details of the standard OGLE
observing techniques, and the DoPhot photometry is available from the OGLE
web site at {\it http://www.astrouw.edu.pl/\~\,ogle/ogle2/ews/ews.html}.

Wo\'zniak et al. (2001) presented a sample of microlensing events from
difference image analysis of the three year OGLE-II bulge data.
His difference image analysis pipeline is designed and tuned for the OGLE bulge data
(Wo\'zniak 2000), and is based on the algorithm from Alard \& Lupton (1998)
and Alard (2000). The difference image analysis pipeline returned a catalog of over 200,000
candidate variable objects, for which only very modest assumptions
have been made about the variability type. This sample was further
searched for objects which somewhere in the light curve
showed no significant variations in a window containing about half of
all photometric points, but had one or two brightening episodes (to allow
for binaries, which often have two peaks).
Wo\'zniak et al. (2001) describes the details of the selection process.
Briefly, a brightening episode is defined as 3 or 4 consecutive points
deviating respectively by 4 or 3 $\sigma$ upwards with respect to the 
baseline flux (determined
from the quiet period). At this point there are no assumptions made
about the shape of the light curve near the event. In all, 4424 light 
curves passed the above criteria. Further cuts requiring the 
light curves to be satisfactorily fit by the standard
microlensing model would strongly discriminate against
the non-standard ones such as parallax events
(see Section~1). To avoid this problem, the most efficient way to 
recover as many of these events as possible, including the interesting 
non-standard ones, is still a visual search.
512 events were found in the course of visual inspection
of all 4424 light curves in the weakly filtered sample,
the largest set of microlensing light curves published so far.
Allowing for slightly larger number of searched fields, this is roughly
a factor of 2 more than discovered by the standard photometric pipeline
from essentially the same data (Udalski et al. 2000). 
In contrast to the standard pipeline, the error distribution from the 
difference image analysis follows a 
Gaussian distribution and the photometric error is reduced by a factor of
2--3. These two properties further increase the chance of seeing departures
from the standard microlensing model.
The error bars were re-calibrated so as to enforce the $\chisq$ per degree
of freedom to be unity for the best-fitting model (see \S 4).

\section{Selection Procedure}

We first fit each microlensing candidate
with the standard single microlens
model which is sufficient to describe most events.
In this model, the (point) source, the lens and the observer are all
assumed to move with constant spatial velocities. The standard light curve,
$A(t)$ is given by (e.g. Paczy\'nski 1986):
\beq \label{amp}
A(t) = {u^2+2 \over u \sqrt{u^2+4}},~~
u(t) \equiv \sqrt{\u0^2 + \tau(t)^2},
\eeq
where $\u0$ is the impact parameter (in units of the Einstein radius) and 
\beq \label{tau}
\tau(t) = {t-t_0 \over \tE}, ~~ \tE = {\tilde{r}_{\rm E} \over \tilde{v}},
\eeq
with $t_0$ being the time of the closest approach (maximum
magnification), $\tilde{r}_{\rm E}$ the Einstein radius projected onto the 
observer plane, $\tilde{v}$ the lens transverse velocity relative to 
the observer-source line of sight, also projected onto the observer
plane, and $\tE$ the Einstein radius crossing time \footnote{In this
paper, we differentiate $\tE$ from the duration of a microlensing event, 
which is defined at the end of this section.}.  The Einstein radius
projected onto the observer plane is given by
\beq \label{rE}
\tilde{r}_{\rm E} = \sqrt{4 G M \Ds x \over {c^2 (1-x)}},
\eeq
where $M$ is the lens mass, $\Ds$ the distance to the source and
$x=\Dl/\Ds$ is the ratio of the distance to the lens and the distance
to the source. Eqs. (\ref{amp}-\ref{rE}) shows the well-known
lens degeneracy, i.e., from a measured $\tE$,
one can not infer $\tilde{v}$, $M$ and $x$ uniquely even if the source distance
is known. 

The flux difference obtained from difference image analysis can be
written as
\beq \label{eq:f}
f(t) = \fl \left[A(t) - 1\right] +\Delta f,
\eeq
where $\fl$ is the baseline flux of the lensed star, and $\Delta f$ is
the difference between the total baseline flux [i.e. the 
flux of the unlensed source and the blended star(s), if present]
and the flux of the reference image.
Note that in general $\Delta f$ does not have to be zero (or even
positive).  Therefore, to fit 
the {\it I}-band data with the standard model, we need five
parameters, namely, $\fl$, $\Delta f$, $\u0, t_0$, and $\tE$.
Best-fitting parameters (and their errors)
are found by minimizing the usual $\chisq$ using the MINUIT program in the
CERN library$\footnote{http://wwwinfo.cern.ch/asd/cernlib/}$.

We then proceeded to fit these light curves with a model which 
accounts for the parallax effect. To do this we need
to describe the lens trajectory in the ecliptic plane. This requires two
further parameters, namely the projected Einstein radius onto the observer plane, 
$\rEt$, and an angle $\psi$ in the ecliptic plane, which is defined 
as the angle between
the heliocentric ecliptic $x$-axis and the normal to the
trajectory
(This geometry is illustrated in Fig.~5 of Soszy\'nski et al. 2001).
Once these two parameters are specified, the resulting lens trajectory in the 
ecliptic plane completely determines the separation between the lens and the 
observer (i.e., the quantity which is analogous to the standard model's $u_0$ 
parameter from eq. 1). This allows the light curve to be
calculated; the complete prescription is given in Soszy\'nski et
al. (2001), to which we refer the reader for further technical 
details (see also Alcock et al. 1995; Dominik 1998).
For the parameters $\fl$, $\Delta f$, $\u0, t_0$, and $\tE$, we take the fit
parameters from the standard fit as the initial guesses, while $\rEt$
and $\psi$ are arbitrarily chosen (see below for more details). The best-fitting
model is again found by minimizing the $\chi^2$.
While the standard fit to an observed microlensing light curve is almost 
always unique, this is not necessarily the case for the parallax events.
To avoid missing the best-fitting parallax fit in the multi-dimensional
parameter space, we ran the parallax fitting program with 24
combinations of $\rEt$ and $\psi$ as initial guesses; the model
with the lowest $\chi^2$ is selected as the best parallax fit.

We then compared the $\chi^2$ values for the standard and 
parallax fits to determine which events were better described by the 
parallax model. Since the parallax fit utilized two additional parameters, 
we had to establish whether a given improvement in $\chi^2$ was simply 
due to the increase in the number of free parameters, or whether this 
improvement was actually due to a deficiency in the standard model.
This was done by performing the standard F-test for
the significance of parameters (e.g. Lupton 1993) on each event. Briefly, we
first calculate the variable 
\beq \label{eq:f-stat}
l = \left[{\chi^2_{\rm S}/{\chi^2_{\rm P}-1}}\right] \times (N-7)/2,
\eeq
where $\chi^2_{\rm S}$ and ${\chi^2_{\rm P} }$ are the $\chi^2$ for the best
standard and parallax fits, respectively, $N-7$ is the number of degrees
of freedom
for the parallax fit, and the factor of 2 is the difference in the degrees
of freedom between the parallax fit and the standard fit. The
variable $l$ follows an $F_{2,n-7}$ distribution. The
statistical significance can then be evaluated using a one-sided
test, i.e., the upper-tail of the $F_{2,n-7}$ distribution. 
A large value of $l$, i.e., a small one-sided probability,
$p_F \equiv F(>l)$, means that there is a  significant improvement 
using the parallax model and vice versa. We selected
events with probability less than $p_F<0.001$ as potential parallax
candidates. In total, 109 microlensing candidates passed this selection
criterion. We visually examined many of these candidates, and found
that the database was still contaminated by events with low 
signal-to-noise ratios. We needed to narrow down the parallax
candidates further and to do this we employed two additional 
criteria: the event `duration' and the peak signal-to-noise ratio.

The duration constraint is imposed because
the parallax effect should, in general, be more prominent for events
that have long duration, since the Earth is able to move a substantial
distance around the Sun during these events.
However, the Einstein radius 
crossing time, $\tE$, can sometimes be misleading.
For example, bright events with large peak fluxes and small photometric error
bars exhibit noticeable magnification for a
period longer than $\sim 2\tE$ because the microlensing variability can 
still be detected even outside the Einstein ring.
To avoid this problem we classified the duration
of an event as being the length of time which the flux of the standard
fit was 3$\sigma$ above its baseline value. This corresponds 
to the length of time which we can practicably utilize in our analysis of the 
lensed part of the light curve. For every 
event we recorded the duration along with the previously mentioned
probability, and events with duration greater than 100 days were 
considered as potential parallax affected candidates. In total, 18
microlensing events passed this additional test. 

However, more than two thirds of these events still turn out
to have noisy data and are unsuitable for detecting 
parallax signatures. A typical example of this is given in Fig. 1(a), which 
clearly illustrates the fact that more accurate data are
needed in order to observe 
the often slight deviations between the standard and parallax fits. Another 
common problem which occurs with the noisy events  is shown in Fig. 1(b).
In this case, the parallax fit substantially improves the $\chi^2$ because the
parallax fit has multiple peaks that better reproduce the
fluctuations in the data. However, one notices that the
secondary peak falls in a gap between two observing seasons. 
Many other parallax fits for the noisy events
show similar multiple-peak structures.
While such multi-peak parallax events have been predicted (Gould et al. 1994),
the large error bars make the identification un-convincing. 
This is further supported by their often unusual parameter values; for example,
the majority of these events have values of $\rEt$ which are significantly
different from the value of a few AU which is expected for a typical 
microlensing event. It is therefore clear that 
the identification of parallax signatures 
requires accurate data in order to observe the slight deviations and 
so the light curves for these noisy events were of little use and we do
not regard them as parallax events.
To eliminate these events from our selection procedure we needed to
quantify how noisy the data were for an event. This was done by analysing 
the signal-to-noise ratio. To assess this signal-to-noise ratio, we use 
a quantity, $\snpeak$, defined as the difference between the 
peak flux and the baseline flux divided by 
the average error estimated using the data points outside the 
`duration' of the light curve. We find empirically that
$\snpeak>30$ provides an excellent separator. This criterion,
while satisfactory, is somewhat arbitrary; we return to this issue briefly
in the discussion.

As well as identifying events which displayed parallax signatures,
we also wished to find events which unexpectedly exhibited no
deviations from the standard light curve. Of particular interest
are events which have unusually large time-scales but no signs of
asymmetry, since they provide lower-limits on $\rEt$ (see \S4.3).
To isolate these we used the event duration which was mentioned
earlier. We also used two further criteria: firstly, a measure of the 
improvement in $\chi^2$ between the standard and the parallax 
fits (i.e. a high value of $p_F$, which corresponds to no significant 
improvement in $\chi^2$) and secondly, the signal-to-noise
ratio, $\snpeak$.
Events which had duration $> 100$
days,  $\snpeak> 30$, and probability $p_F> 0.05$ were considered as
long-duration events that showed no parallax signatures.

\section{Results}

Out of the 512 events, there were 8 events where the incorporation of
the parallax effect substantially improved the goodness of the fit (see
above) and passed our test of event duration and the peak
signal-to-noise ratio. These events are given in Table 1. 
One light curve was found 
to show clear signs of parallax affected behaviour, in addition to
a number of others which could be classified as marginal cases.
In the following, we shall first discuss these parallax candidates and
in \S4.3 we discuss the physical limits that one can
derive from the long-duration events that exhibited no parallax signatures.

For all the microlensing events
presented here, we re-normalize the $\chi^2$ per degree of freedom to
be unity for the best-fitting model by multiplying the quoted
observational errors by a constant factor (usually between 0.8 to 1.3).
This is necessary because we found that the quoted error bars in
observations are often too large. The events in \S 4.1 and \S 4.2 
were re-normalized using the best-fitting parallax model, while the events
in \S 4.3 were re-normalized using the best-fitting standard model (which has
essentially the same $\chi^2$ as the best-fitting parallax model).
This re-normalization affects 
primarily the quoted error bars 
in the model parameters. It also weakly affects the selection
procedure through both the peak flux ratio, $\snpeak$,
and duration, although this perturbation is only very slight 
since $l$ in eq. (\ref{eq:f-stat}) remains unchanged.

\subsection{Parallax microlensing candidate}

  From the catalogue of 512 microlensing events, the most convincing
parallax event we found was sc33\_4505, which is shown in Fig. 2.
Despite the lack of data for two important regions
(namely $t \equiv {\rm JD}-2450000<550$ days and $760<t< 870$ days), 
the systematic deviations from the standard curve can clearly be seen.
During the upward slope of the light curve the standard fit undoubtedly
has a gradient which is too shallow and yet for the downward slope
this fit's gradient is clearly too steep. This prominent asymmetry
provides strong evidence that the light curve may be exhibiting
parallax behaviour and this is further supported when the parallax
model is used to fit the data. This model provides a much more accurate
fit, with the $\chi^2$ being drastically reduced from 398.972 (or $\chi^2=2.23$
per degree of freedom) to 177 ($\chi^2=1$ per degree of freedom).
When observing resumed after $t = 870$ days it can be seen that 
the parallax model fitted this region more accurately, even though
the lensing event was very nearly over and the magnification was
only a fraction of the peak value. 

The best fit parameters from both the standard and parallax fit are 
given in Table 2.  In particular, for the best-fitting parallax model, we find that
\beq
\tE = 194\pm 30\,{\rm days},  ~~ \rEt = 6.37\pm 0.53\,{\rm AU}.
\eeq
An expression for the lens mass as a function of the relative
lens-source distance can be shown to be (see Soszy\'nski et al. 2001),
\beq \label{eq:limit}
M = \frac{c^2 \tilde{r}_{\rm E}^2}{4G} \left( {{1 \over \Dl}
- {1  \over \Ds} } \right)
= 0.0123 M_\odot \left( \frac{\tilde{r}_{\rm E}}{\rm AU} \right)^2
\left( \frac{\pirel}{0.1 \rm mas} \right),
~~\pirel\equiv{{{\rm AU} \over \Dl} - {{\rm AU} \over \Ds} },
\eeq
where $\pirel$ is the relative lens-source parallax in milli-arcseconds.
For this event, the parallax parameters give a projected velocity of
\beq
\tilde{v} = \frac{\tilde{r}_{\rm E}}{\tE} = 57^{+16}_{-12} \mbox{ km s}^{-1}.
\eeq
The lens mass depends on the relative lens-source parallax
(eq. \ref{eq:limit}). If the source is about 7\,kpc away, and
the lens lies in the disk half-way between the observer and the 
source ($x=1/2$),
then $\pirel \approx 0.143$ mas, which gives a lens mass of about 
$0.71M_\odot$; as a comparison, for a bulge self-lensing event with 
$\Ds \approx 8\,\rm{kpc}$ and $\Dl \approx 6\, \rm{kpc}$,
then $\pirel \approx 0.042$ mas, which would give a lens mass of
about $0.21M_\odot$.
However, this latter scenario is less likely since the projected 
velocity of the lens is relatively low.
It appears that similar to the previous parallax microlensing events,
this event is caused by a slow-moving and likely low-mass lens.

\subsection{Marginal parallax microlensing candidates}

Apart from the above convincing candidate, sc33\_4505, a number of events were
found which showed indications of parallax signatures. However, the
case for classifying these events as having parallax affected light curves
is not wholly convincing, mainly due to the insufficiency of the data.
Figs. 3-9 show these `marginal' microlensing events that we have found,
and Table 3 presents the best fit parameters for both the standard and 
parallax models. 

 The problem of insufficient sampling is clear from
several marginal cases. For example, for the sc41\_3299 event (Fig. 8),
there are very few points in the rising branch of the light
curve. As a result, although the parallax fit seems to provide a somewhat
better fit in the region between $t=880-900$ days, it is not clear  
whether this is a genuine parallax event or just due to the large
errors in the data. The situation is similar for events sc20\_5748 (Fig. 4),
sc27\_3078 (Fig. 6), and sc43\_836 (Fig. 9). For the event sc6\_2563
(Fig. 3), the standard 
fit consistently over-predicts the fluxes between $t=1040$ to $1140$ days, and the
parallax model provides a much better fit for these parts. Unfortunately,
there was no data for the time period ($t\sim$ 1180 days)
when the standard model and the
parallax model are predicted to show substantial deviations.
The fact that we can pick out marginal cases like sc41\_3299 implies that
our selection method is sensitive, and it is unlikely that we
could have missed many true candidate parallax events.

The situations for the events sc20\_5875 (Fig. 5) and
sc35\_2526 (Fig. 7) are slightly different. For the 
sc35\_2526 event, the light curve is highly asymmetric, and the standard 
model is clearly a bad fit. The parallax model, on the other hand,
provides a much better fit; the improvement in $\chi^2$
is from 361.5 to 170. However, the fit is not
perfect, particularly at the peak and for the few data points between
$700<t<750$ days. For this reason, we still cautiously regard this event 
as a ``marginal'' parallax candidate event. The situation for sc20\_5875 
is similar, but less dramatic. It is possible that these two
events are produced by binaries, where the asymmetry is provided, e.g.,
by the shear perturbation by the secondary on the primary lens
(e.g. Mao \& Di Stefano 1995). Notice that, for the event sc35\_2526, the
$\tE$ parameter is unusually long (see Table 3); however, this is just an artifact 
due to the well-known degeneracy between the impact parameter, $u_0$, and $\tE$ for
microlensing light curves in blended light curves (Wo\'zniak \&
Paczy\'nski 1997), where only the combination of $u_0\tE$ can be inferred
in some cases.

\subsection{Long microlensing events that show no parallax signatures}

Of almost equal interest to parallax events are the events
which have large time-scales and
 yet exhibit no signs of parallax induced asymmetry. A total of 20 events
were identified using the selection criteria which were given in \S3, i.e. 
events which had duration $> 100$ days, $\snpeak>30$, 
and probability $ p_F> 0.05$. The parallax signatures become more
pronounced for small values of $\rEt$ because the Earth's motion around
the Sun then becomes a larger fraction of the Einstein radius, and hence
presents a larger perturbation to the light curve. The lack of parallax
signatures therefore provides a lower-limit on $\rEt$. For each such
event, we fix $\rEt$ in a range of values 
while letting all the other six parameters vary and find the minimum
$\chi^2$ value for each fixed $\rEt$. The $2\sigma$ confidence constraint
on $\rEt$ is then given by the $\rEt$ value corresponding to the point
 at which its $\chi^2$ becomes larger than the best fit $\chi^2$ 
by 4.00 (e.g. Lupton 1993). From eq. (\ref{eq:limit}), one sees that
 this lower limit on $\rEt$ can then 
be translated into a lower limit on $M/\pirel$.
The events which produced the 4 largest constraints are given in Table 4,
 and their light curves are shown in Fig. 10.

In all cases, the lower limits on the mass are rather weak, the best
case being for event sc26\_2218, where the lower limit reaches about
$0.2M_\odot$ for $\pirel = 0.1$ mas. These weak constraints are 
similar to those found 
by the EROS collaboration for the microlensing event EROS2-GSA1
(Derue et al. 1999), from which they obtained values of $\rEt \ge 1.33$ AU 
and $M \ga  10^{-3}M_\odot$.

\section{Summary and discussion}

We have conducted an extensive search for parallax events in the microlensing
database of 512 microlensing events obtained by
the difference image analysis of the OGLE II data. Our selection
procedure involves a direct comparison of the standard microlensing 
model and the parallax model, augmented by the requirements of long
duration and high peak signal-to-noise ratio. While empirically we
found these criteria to be very effective, there is no guarantee that these 
selection procedures are optimal. This issue is best explored
using the recovery of artificial parallax events in Monte Carlo simulations
(see below).

Using our selection criteria, we
found one convincing new parallax microlensing event, sc33\_4505. 
A number of other more marginal cases were also
recovered. However, while these marginal events are better fitted by a model
incorporating the parallax effect, their parallax nature
cannot be established beyond any doubt due to the poor sampling in the
light curve. This may be particularly severe for the events which last
for about one year, due to the gap of approximately three months
between observing seasons. In such cases either the rising or the
declining branch of the light curve will typically be missing, and so the
parallax induced asymmetry is particularly difficult to identify.
We have also found some very 
long events that showed no parallax signatures, and these events provide a
lower-limit on the Einstein radius projected onto the  observer
plane. Consequently, a lower-limit on the lens mass $M/\pirel$ can be derived
(cf. eq. \ref{eq:limit}). However, in most cases, the lower limits 
on $M/\pirel$ are rather weak.

The convincing parallax event (sc33\_4505) is similar to all of the
other 4 known parallax events (see introduction) in that it is 
caused by a slow-moving lens and, quite likely, a 
low mass lens. The low projected velocity favors the interpretation that 
this microlensing geometry is a disk source lensed by a disk lens. For such
events, the observer, the lens and the source rotate about the Galactic
center with roughly the same velocity, and the relative motion is only
due to the small, $\sim 10\kms$, random velocities 
(see, e.g. Derue et al. 1999).
On the other hand, the chance for a bulge source (with its much larger
random velocity, $\sim 100\kms$) 
to have such a low projected velocity relative to
the lens (whether in the disk or bulge) is small. So it appears that all
known parallax events are caused by disk-disk lensing. If this is true,
then the radial velocities of the lensed sources are expected to be small.
It will therefore be very interesting to check this by obtaining the
radial velocities of the parallax events spectroscopically.
Since their projected transverse velocities are known, one obtains a
more complete kinematic picture of these unusual
microlensing events that can be used to probe the dynamical model of the
Milky Way. As a by product, one also obtains the metallicity
and age of these stars.

While the number of convincing parallax events toward
the Galactic bulge seems to be very low (1
out of 512), the number of marginal cases makes the true fraction
somewhat uncertain. A related question is whether existing microlensing
catalogs are biased against parallax events. This is an important issue
because parallax events preferentially have long durations, which may be
detectable only when one has monitored the stars for many years. In
fact, after the completion of this work, eight additional events were
recovered by cross identification of the 214 microlensing candidates
from the standard OGLE database with the difference image analysis 
variability database of Wo\'zniak (2000). Some of these were 
missed in the first search because they have such long
durations that they have not yet reached a constant baseline.
We plan to analyze these additional unique events for parallax signatures.
On the theoretical side,
Buchalter \& Kamionkowski (1997) have estimated
the expected fraction of parallax events which would be identified, using 
a regular sampling of every few minutes to $\sim 1$ day.
Our search strongly suggests that the irregular sampling and gaps between 
different observing seasons hamper the recovery of parallax events.
We plan to perform a simulation with realistic sampling and photometric
errors similar to those found in observations. A comparison between the
observed and predicted rates may enable us to provide constraints 
on the lens and source kinematics. Such simulations will
also be helpful for devising the optimal search strategy for selecting
parallax events from the experiments. The results of these Monte Carlo
simulations will be reported in a future publication.

\section*{Acknowledgement}

We acknowledge Bohdan Paczy\'nski for 
discussions and comments on the manuscript. We also thank Ian Browne for
critical remarks that improved the paper. 
MCS acknowledges receipt of a PPARC grant.
PW was supported by the NSF grant AST-9820314
to Bohdan Paczynski and by the Laboratory Directed Research \& Development
funds (X1EM and XARF programs at LANL).

{}

\clearpage

\begin{figure}
\plotone{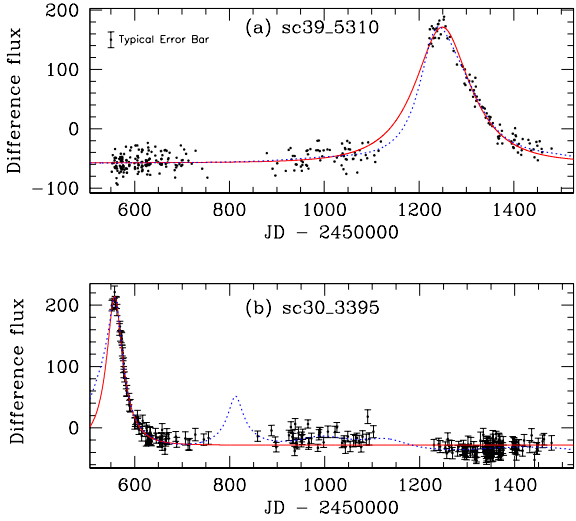}
\caption{Two events which highlight the problems associated with noisy data. 
The solid and dotted lines represent the standard fit and the parallax
fit, respectively.
(a) The event sc39\_5310, which illustrates how slight deviations between the
two fits can easily be hidden by noisy data. The error bars have been 
omitted for clarity, but a 
typical error bar is shown in the top left hand corner.
(b) The event sc30\_3395, which demonstrates how
a microlensing light-curve can be seemingly-erroneously
fitted with a parallax model exhibiting many peaks. This
leads to a significant improvement in $\chi^2$.
Such multi-peak parallax models often occur in events with noisy data,
 which may signal additional variability other than microlensing.
}
\end{figure}


\begin{figure}
\plotone{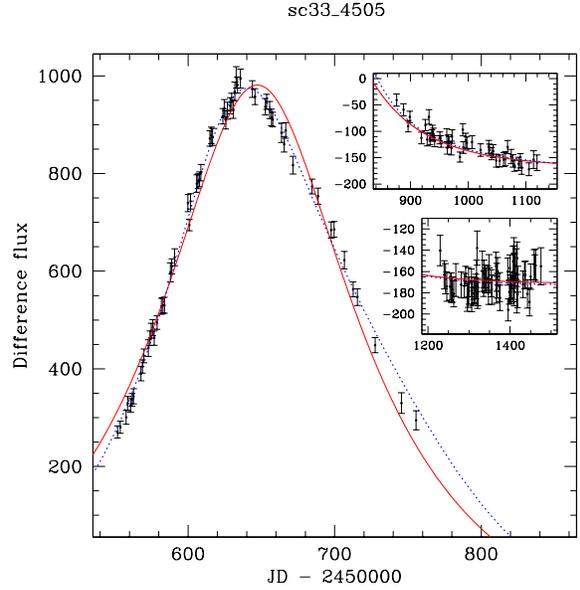}
\caption{Difference flux ($f(t)$, as defined in eq. 4) as a function
of time for the best candidate parallax microlensing event,
sc33\_4505. The solid line shows
the best standard fit while the dotted line shows the best fit that
accounts for the parallax effect. The insets show the light curves for
the `non-lensed' seasons.
}
\end{figure}

\begin{figure}
\plotone{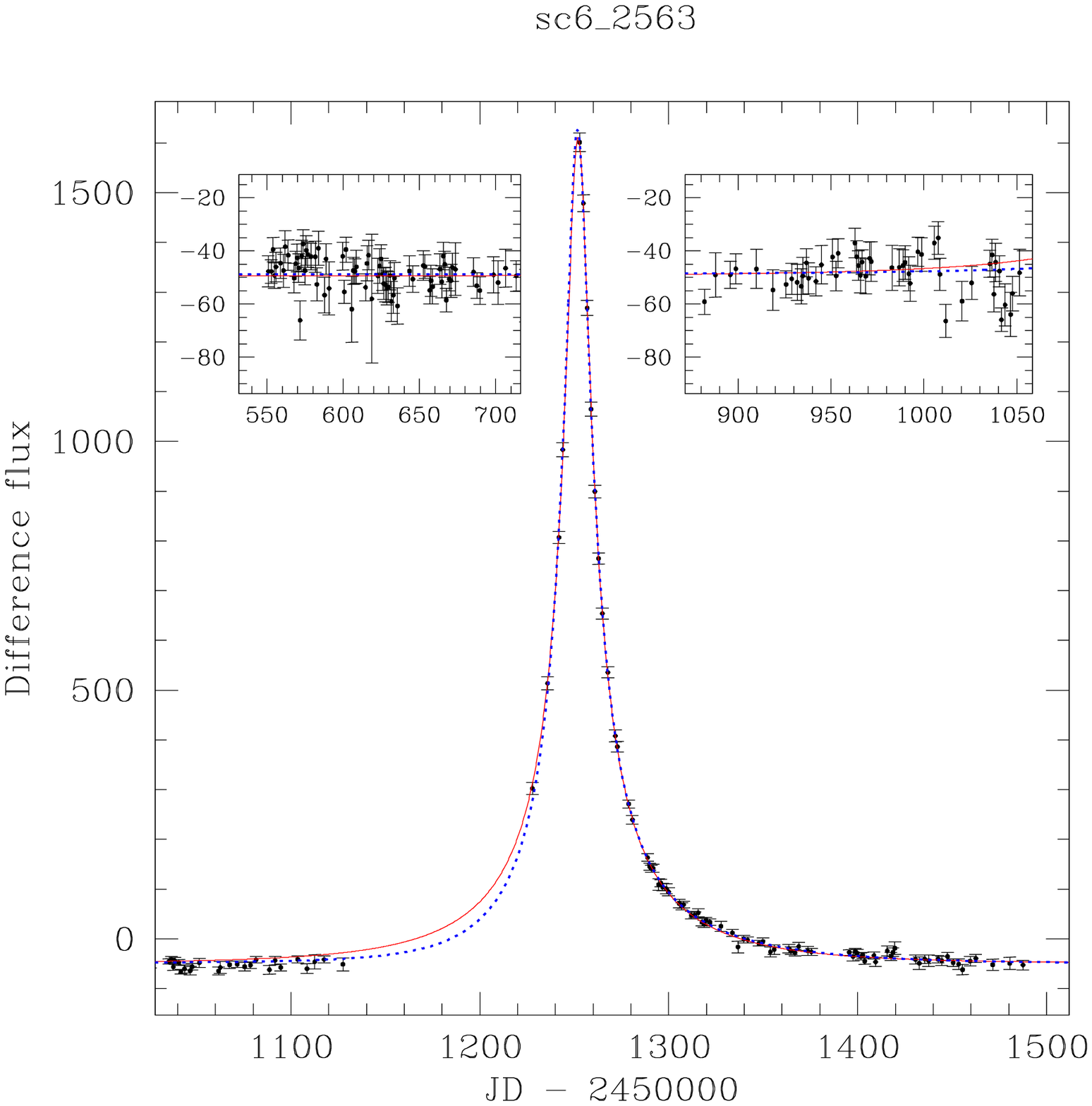}
\caption{The marginal candidate parallax microlensing event,
sc6\_2563.  The notations are the same as in Fig. 2.}
\end{figure}

\begin{figure}
\plotone{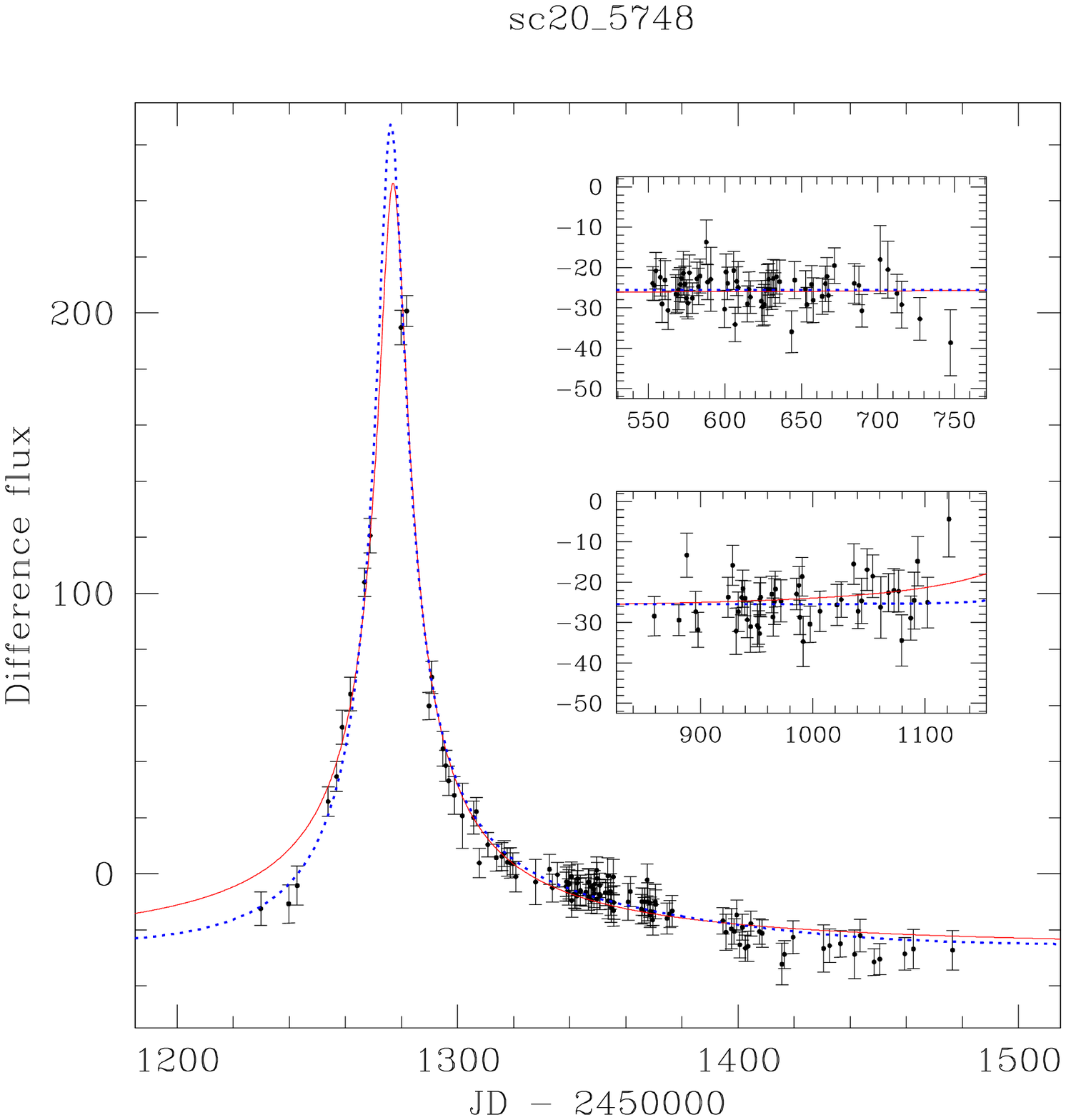}
\caption{The marginal candidate parallax microlensing event,
sc20\_5748. The notations are the same as in Fig. 2.}
\end{figure}

\begin{figure}
\plotone{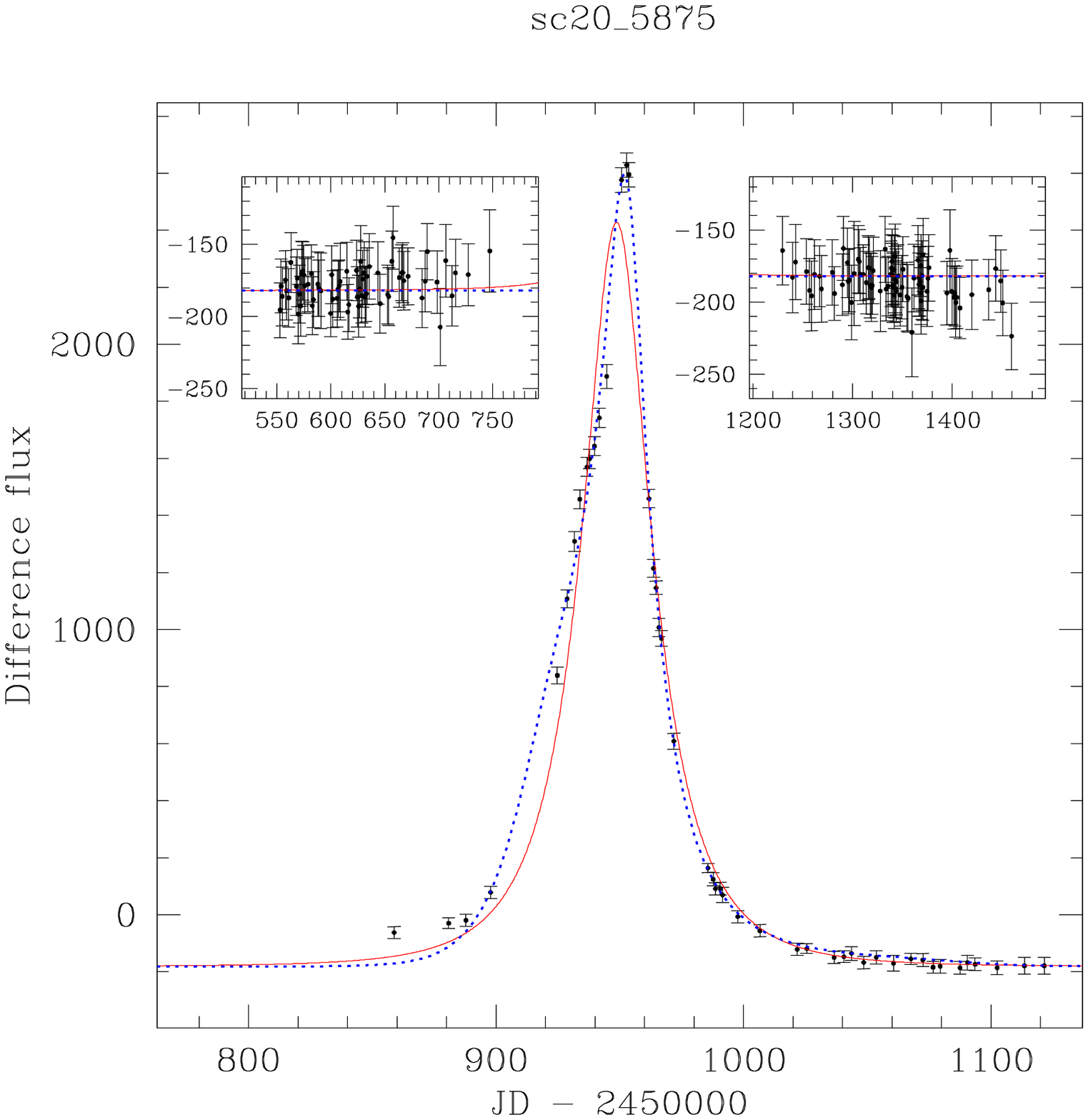}
\caption{The marginal candidate parallax microlensing event,
sc20\_5875.  The notations are the same as in Fig. 2. This event is not
perfectly fit by the parallax model. This
event may, instead, be caused by a (weak) binary lens (see also Fig.\,7). }
\end{figure}

\begin{figure}
\plotone{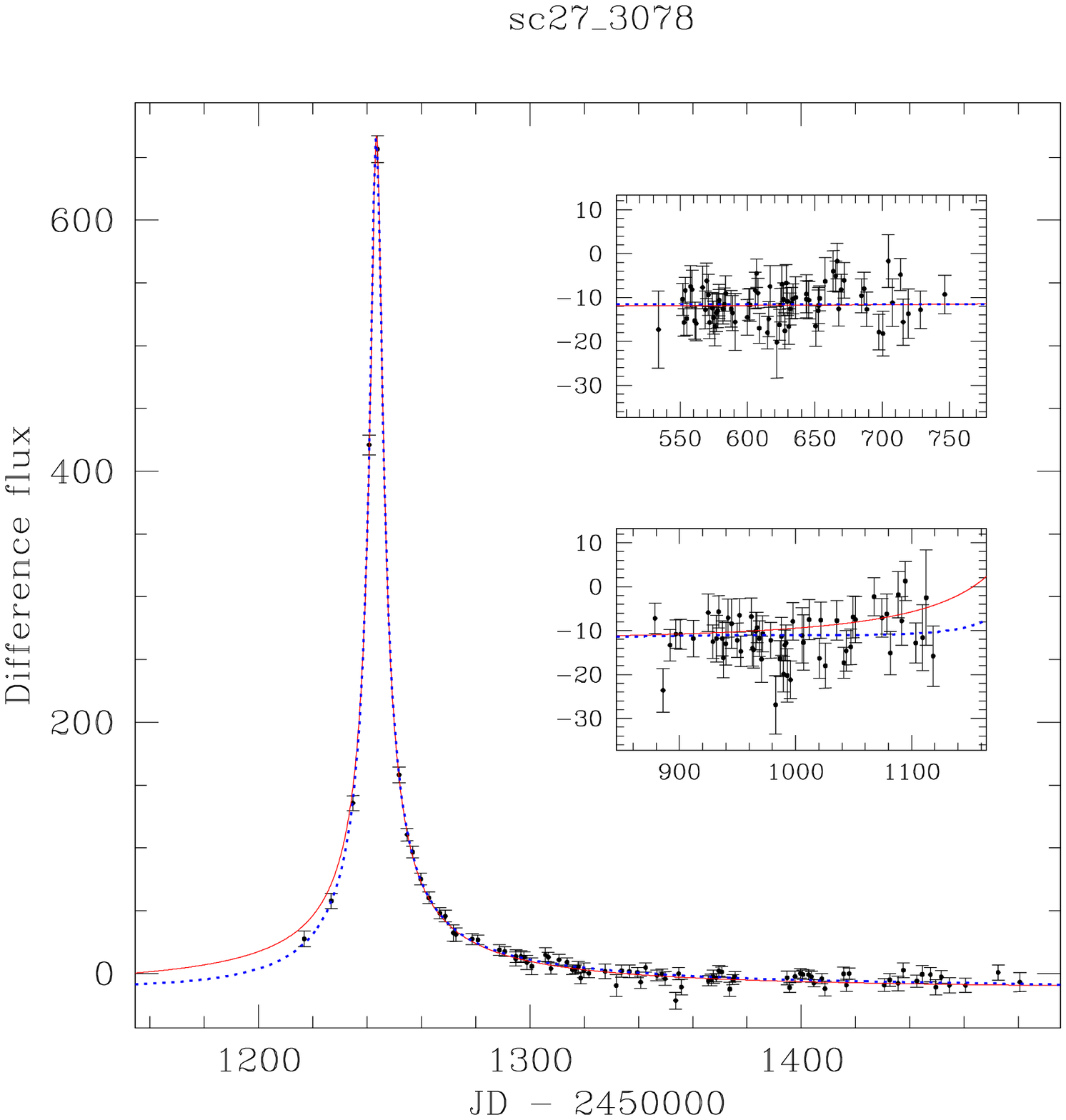}
\caption{The marginal candidate parallax microlensing event,
sc27\_3078.  The notations are the same as in Fig. 2.}
\end{figure}

\begin{figure}
\plotone{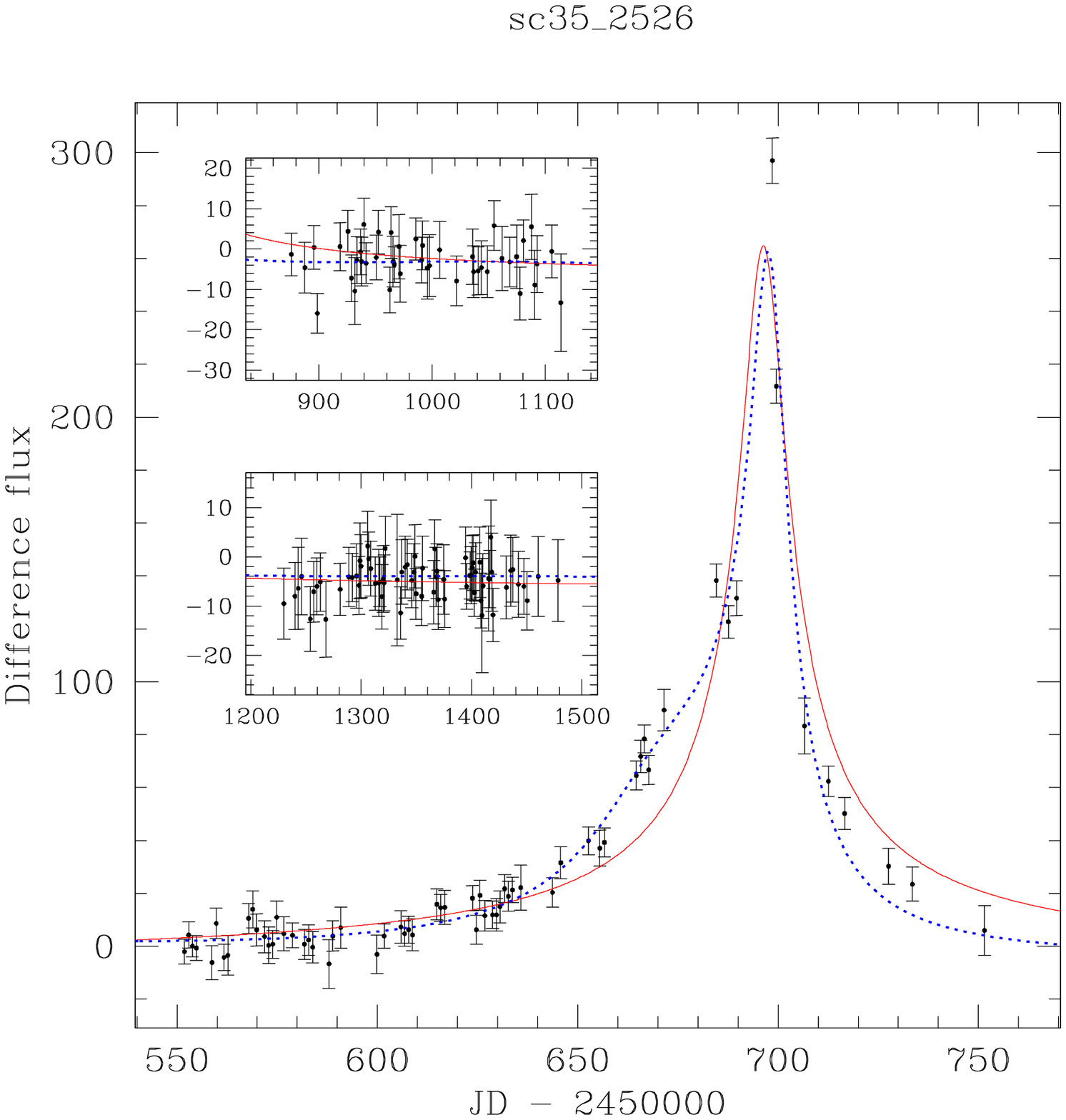}
\caption{The marginal candidate parallax microlensing event,
sc35\_2526.  The notations are the same as in Fig. 2.
This event is not perfectly fit by the parallax model. This
event may, instead, be caused by a (weak) binary lens (see also Fig.\,5).
}
\end{figure}

\begin{figure}
\plotone{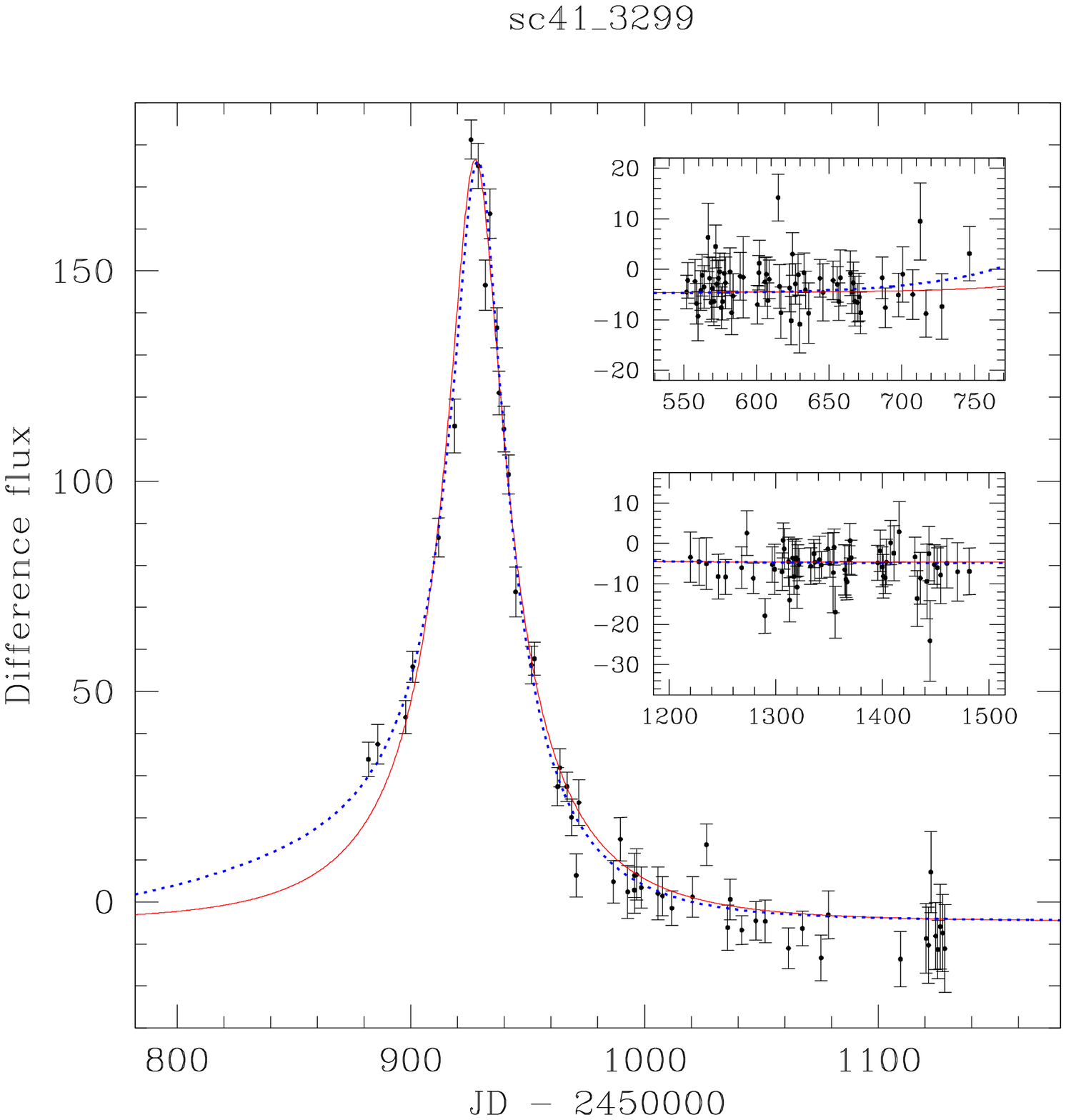}
\caption{The marginal candidate parallax microlensing event,
sc41\_3299. The notations are the same as in Fig. 2.}
\end{figure}

\begin{figure}
\plotone{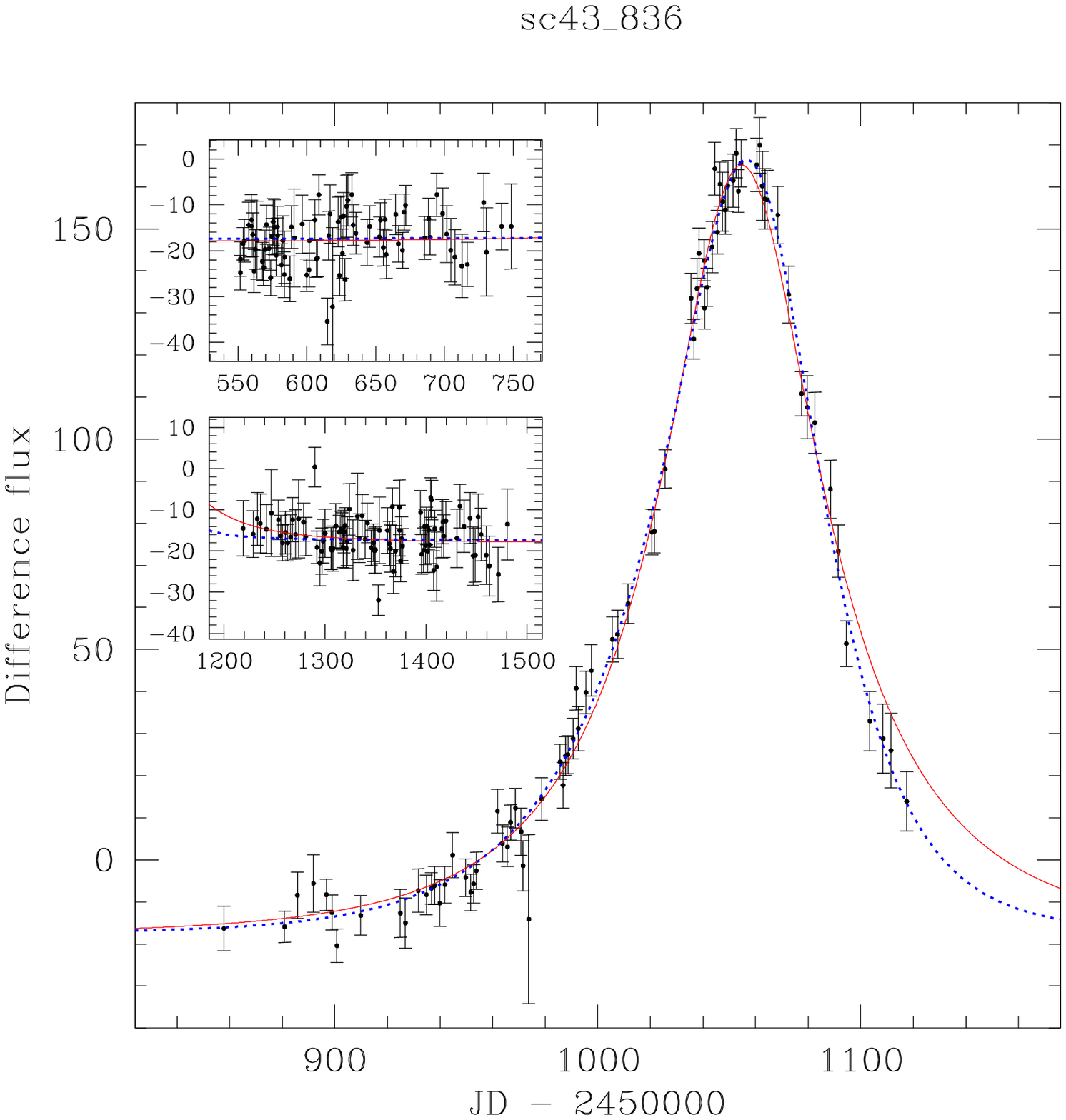}
\caption{The marginal candidate parallax microlensing event, sc43\_836. The notations are the same as in Fig. 2.}
\end{figure}

\begin{figure}
\plotone{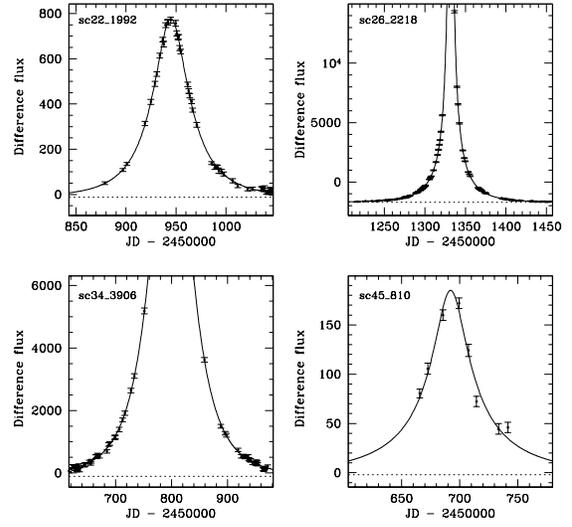}
\caption{The 4 long-duration lensing events which produced the most stringent lower-limits
on the lens mass $M/\pirel$ (see eq. \ref{eq:limit} and Table 4). For these events, the
incorporation of parallax effects produces virtually no improvement in the $\chi^2$.
The dotted line shows  the theoretical baseline flux and
 the horizontal axis covers a length of time corresponding to the event duration, as specified in \S3.}
\end{figure}

\clearpage

\begin{table}
\begin{center}
\caption{ The 8 events which were systematically selected as potential
parallax affected light-curves, and their subsequent classification. The
probability $p_F$ and $\snpeak$ are defined in \S3. The
convincing candidate (sc33\_4505) and other
marginal candidates are shown in Figs. 2-9.}
\tiny
\vspace{0.3cm}
\begin{tabular}{ccccccccc}

Event & RA & DEC & Duration (day) & $\tE$ (day) & $p_F$ & {$\snpeak$} 
& Classification \\
\hline
%
%
%
%
%
%
sc6\_2563 & 18:08:02.64 & -31:49:05.2 &
 $ 245 $ & $ 66.6\pm 1.3 $ & $ 1.22 \times 10^{-8} $ & 228.39 & marginal \\
%
%
%
sc20\_5748 & 17:59:39.05 & -28:27:16.6 &
 $ 150 $ & $ 78.2\pm 2.5 $ & $ 2.58 \times 10^{-5} $ & 38.99 & marginal  \\
sc20\_5875 & 17:59:08.99 & -28:24:54.7 &
 $ 146 $ & $ 27.22\pm 0.21 $ & $ 0.00 $ & 119.04 & marginal/binary?
\\
%
%
%
sc27\_3078 & 17:48:15.92 & -34:50:09.0 &
 $ 162 $ & $ 124\pm 10 $ & $ 2.04 \times 10^{-7} $ & 142.82 & marginal 
\\
%
%
%
%
sc33\_4505 & 18:05:46.71 & -28:25:32.1 &
 $ 663 $ & $ 194\pm 30 $ & $ 5.78 \times 10^{-32} $ & 78.79 & convincing \\
%
%
%
sc35\_2526 & 18:04:22.42 & -27:57:52.2 &
 $ 162 $ & $ 6600\pm 180 $ & $ 1.44 \times 10^{-28} $ & 47.02 & marginal/binary?
\\
%
%
%
sc41\_3299 &17:52:19.08 & -32:48:20.8 &
 $ 119 $ & $ 98\pm 18 $ & $ 1.28 \times 10^{-5} $ & 35.17 & marginal 
\\
%
%
sc43\_836 &17:34:52.64 & -27:23:14.4 &
 $ 207 $ & $ 45.1\pm 1.6 $ & $ 4.31 \times 10^{-5} $ & 31.70 & marginal
\\ 
%
%

\end{tabular}
\end{center}
\end{table}

\begin{table}
\begin{center}
\tiny
\caption{The best standard model (first row) and the best
parallax model (second row) for the 
convincing candidate sc33\_4505.
} 
\vspace{0.3cm}
\begin{tabular}{ccccccccc}
Model & $t_0$ & $\tE$ (day) & $\u0$ & $\fl$
& $\Delta f$ & $\psi$ & $\rEt$ (AU) & $\chisq$  \\
\hline
S & $ 647.22\pm       0.36$ &
$    166\pm      11$ &
$     -0.412\pm       0.036$ &
$    730\pm      91$ &
$   -172.2\pm       1.8$ &
--- & --- & 399.0
\\
P & $ 654.4\pm       4.1$ &
$    194\pm      30$ &
$     -0.207\pm       0.065$ &
$    440\pm     150$ &
$   -174.3\pm       2.1$ &
$      3.136\pm       0.011$ &
$      6.37\pm       0.53$ & 177
\\

\end{tabular}
\end{center}
\end{table}

\begin{table}
\begin{center}
\caption{The best standard model (indicated by `S') and the best
parallax model (indicated by `P') 
for marginal parallax candidates.
The error bars have been rescaled so as to enforce the $\chi^2$ per
degree of freedom to be unity for the best-fitting parallax model.
}
\tiny
\vspace{0.3cm}
\begin{tabular}{ccccccccc}

Event & $t_0$ & $\tE$ (day) & $\u0$ & $\fl$
& $ \Delta f$ & $\psi$ & $\rEt$ (AU) & $\chisq$  \\
\hline
sc6\_2563,S &
$   1251.971\pm       0.068$ &
$     89.2\pm       2.8$ &
$      0.0758\pm       0.0032$ &
$    135.9\pm       5.7$ &
$    -49.43\pm       0.52$ &
--- & --- & 253.7
\\
------------,P &
$   1263.85\pm       0.62$ &
$     66.6\pm       1.3$ &
$      0.0272\pm       0.0085$ &
$    192.8\pm       5.5$ &
$    -48.84\pm       0.49$ &
$      0.262\pm       0.012$ &
$      4.27\pm       0.16$ & 214
\\
sc20\_5748,S &
$   1277.03\pm       0.22$ &
$    357\pm     120$ &
$     -0.0148\pm       0.0056$ &
$      4.1\pm       1.5$ &
$    -26.24\pm       0.67$ &
--- & --- & 243.2
\\
------------,P &
$   1315.12\pm       0.96$ &
$     78.2\pm       2.5$ &
$     -0.325\pm       0.041$ &
$     16.67\pm       0.79$ &
$    -25.50\pm       0.41$ &
$      1.234\pm       0.057$ &
$      2.43\pm       0.11$ & 221
\\
sc20\_5875,S &
$    948.74\pm       0.10$ &
$     31.3\pm       1.7$ &
$      0.461\pm       0.042$ &
$   1950\pm     250$ &
$   -181.9\pm       1.7$ &
--- & --- & 666.7
\\
------------,P &
$    933.88\pm       0.28$ &
$     27.22\pm       0.21$ &
$      2.073\pm       0.011$ &
$    320\pm      25$ &
$   -182.0\pm       1.6$ &
$      0.235\pm       0.011$ &
$      0.3515\pm       0.0021$ & 189
\\
sc27\_3078,S &
$   1243.481\pm       0.077$ &
$    312\pm      77$ &
$      0.0070\pm       0.0018$ &
$      4.8\pm       1.2$ &
$    -12.08\pm       0.52$ &
--- & --- & 245.2
\\
------------,P &
$   1238.75\pm       0.34$ &
$    124\pm      10$ &
$     -0.6266\pm       0.0068$ &
$     18.2\pm       1.3$ &
$    -11.60\pm       0.36$ &
$     2.971\pm       0.026$ &
$      1.68\pm       0.019$ & 212
\\
sc35\_2526,S &
$    696.33\pm       0.20$ &
$  53000\pm   11000$ &
$      0.000107\pm       0.000021$ &
$      0.0291\pm       0.0057$ &
$     -7.39\pm       0.59$ &
--- & --- & 361.5
\\
------------,P &
$    695.46\pm       0.23$ &
$   6600\pm     180$ &
$      0.01512\pm       0.00018$ &
$      0.0920\pm       0.0066$ &
$     -4.65\pm       0.57$ &
$      2.9967\pm       0.0087$ &
$     63.02\pm       0.73$ & 170
\\
sc41\_3299,S &
$    927.65\pm       0.29$ &
$     63.2\pm       8.3$ &
$      0.186\pm       0.034$ &
$     40.8\pm       8.5$ &
$     -4.63\pm       0.40$ &
--- & --- & 197.1
\\
------------,P &
$    913.1\pm       7.7$ &
$     98\pm      18$ &
$      0.19\pm       0.16$ &
$     26.2\pm       6.8$ &
$     -4.87\pm       0.47$ &
$     6.222\pm       0.096$ &
$      2.33\pm       0.68$ & 173
\\
sc43\_836,S &
$   1054.69\pm       0.41$ &
$     71.5\pm       7.2$ &
$      0.437\pm       0.067$ &
$    126\pm      28$ &
$    -17.91\pm       0.41$ &
--- & --- & 262.0
\\
------------,P &
$   1040.0\pm       1.6$ &
$     45.1\pm       1.6$ &
$     -0.104\pm       0.035$ &
$    176\pm      13$ &
$    -17.38\pm       0.37$ &
$      3.1744\pm       0.0054$ &
$      1.511\pm       0.091$ & 241
\\

\end{tabular}
\end{center}
\end{table}

\begin{table}
\begin{center}
\caption{The 4 non-lensed events which produced the best $2\sigma$
confidence lower limit on the projected Einstein radius in the observer
plane ($\rEt$) and on $M/\pirel$ (cf. eq. \ref{eq:limit}). The
`duration' of the event is defined in \S3, and
the probability $p_F$ is defined below eq. (\ref{eq:f-stat}).
The relative lens-source parallax $\pirel$ is in units of
milli-arcseconds; for microlensing events toward the Galactic bulge, 
$\pirel \sim 0.1$ mas. 
} 
\tiny
\vspace{0.3cm}
\begin{tabular}{cccccccc}
Event & RA & DEC & Duration (day) &  $\tE$ (day) & $p_F$ & $\tilde{r}_{\rm E}$
(AU) & $M/\pirel$ ($M_{\odot}$) \\
\hline
sc22\_1992& 17:56:35.30 & -30:56:33.4 & $151$ & $45.5\pm 3.7$ & $0.132$ & $0.94$ & $0.11$ \\
sc26\_2218&17:47:23.29 & -34:59:52.4 & $195$ & $45.10\pm 0.59$ & $0.104$ & $3.74$ & $1.71$ \\
sc34\_3906&17:58:37.11 & -29:06:29.9& $331$ & $58.7\pm 6.2$ & $0.204$ & $1.44$ & $0.25$ \\
sc45\_810& 18:03:30.06 & -30:09:55.6 & $115$ & $98\pm 54$ & $0.151$ & $1.73$ & $0.37$ \\
\end{tabular}
\end{center}
\end{table}

\end{document}